\documentstyle[aps,prc,epsfig]{revtex}

\begin{document}


\title{Final State Interaction in Exclusive $(e,e'NN)$ Reactions}

\author{D.\ Kn\"odler and H.\ M\"uther}

\address{Institut f\"ur Theoretische Physik, Universit\"at T\"ubingen,\\
Auf der Morgenstelle 14, D-72076 T\"ubingen, Germany  }


\maketitle

\begin{abstract}
Contributions of nucleon-nucleon (NN) correlations, meson exchange currents and
the residual final state interactions (FSI) on exclusive two-nucleon knock-out
reactions induced by electron scattering are investigated.  All contributions
are derived from the same realistic meson exchange model for the NN
interaction.  Effects of correlations and FSI are determined in a consistent
way by solving the NN scattering equation, the Bethe-Goldstone equation, for
two nucleons in nuclear matter. One finds that the FSI re-scattering terms are
non-negligible even if the two nucleons are emitted back to back. 
\end{abstract}


\section{Introduction}
Exclusive $(e,e'NN)$ reactions, as well as photoinduced two nucleon knock-out
experiments ($\gamma,NN$) experiments, are often considered as very powerful
tools to explore correlations in the nuclear many-body wave function, which
are beyond the mean field or Hartree-Fock approximation. Using a simple picture
for such reactions the real or virtual photon of inelastic electron scattering 
is absorbed by a pair of nucleons. This pair of nucleons is knocked out from the
target nucleus while the residual (A-2) nucleons can be considered as spectators
which form the ground state or another well defined bound state of the
daughter nucleus. Due to the progress in accelerator and detector technology 
such triple coincidence experiments with sufficient resolution to identify
specific states of the daughter nucleus have become possible and first
results have been reported in the literature\cite{blom,onder,rosner,grab1}.

If the mean field approximation, in  which the nucleons move independent from
each other in the Hartree-Fock field, would be valid for the nuclear wave
function, such processes should be strongly suppressed. Therefore the analysis
of these reactions may provide detailed information about the correlations
between the two nucleons which absorb the photon. Various modern models for the
nucleon-nucleon (NN) interaction, which all yield an excellent fit to the NN
scattering data below the threshold for pion production\cite{cdb,argv18,nijm1},
provide different predictions for the short-range and tensor
correlations\cite{pol1,pol2}. Therefore one hopes that the analysis of exclusive
two-nucleon knock-out experiments will provide additional constraints on the
models for the NN interaction.

\begin{figure}[t]
\begin{center}
\epsfig{file=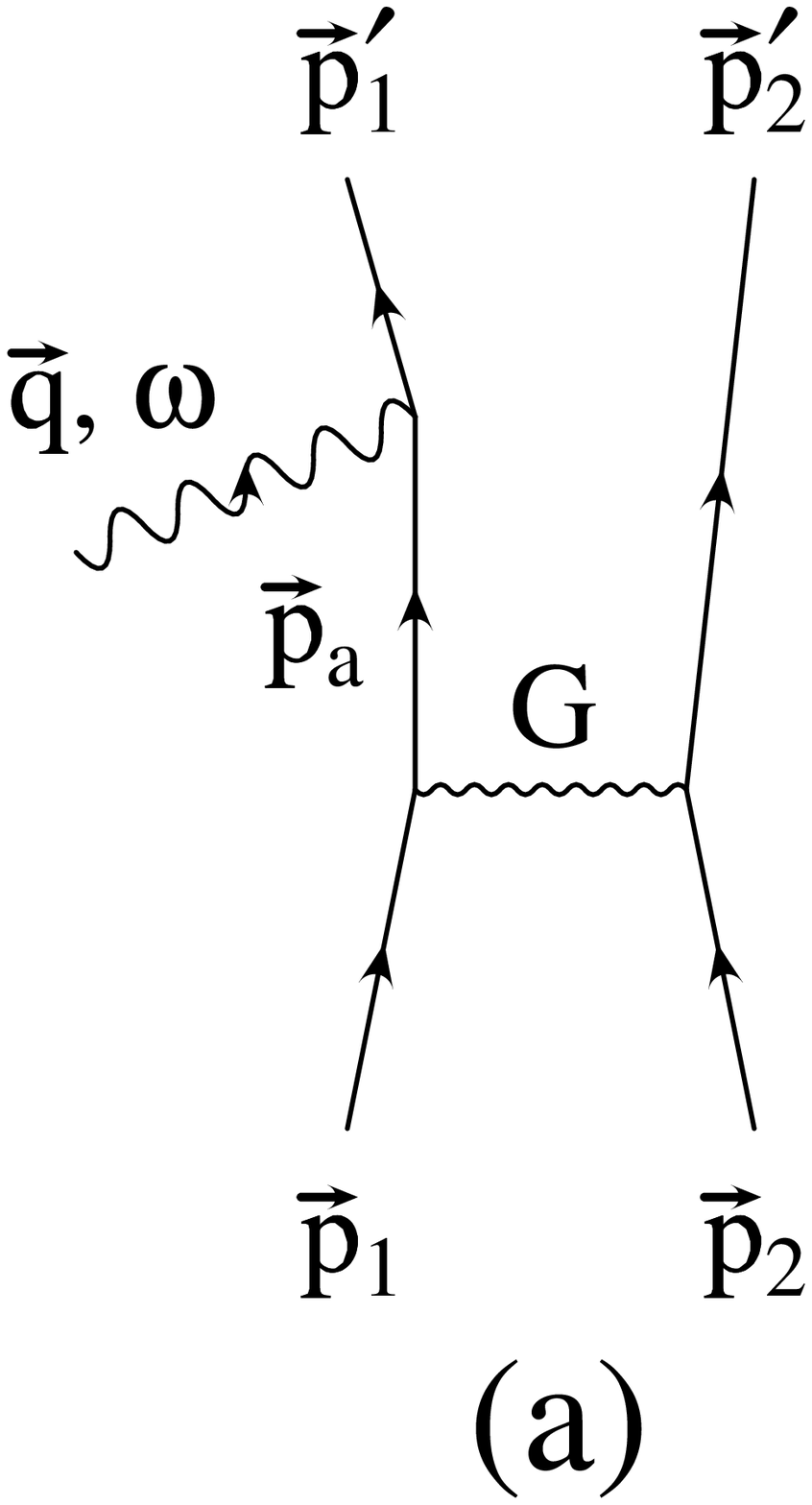,scale=0.25}
\epsfig{file=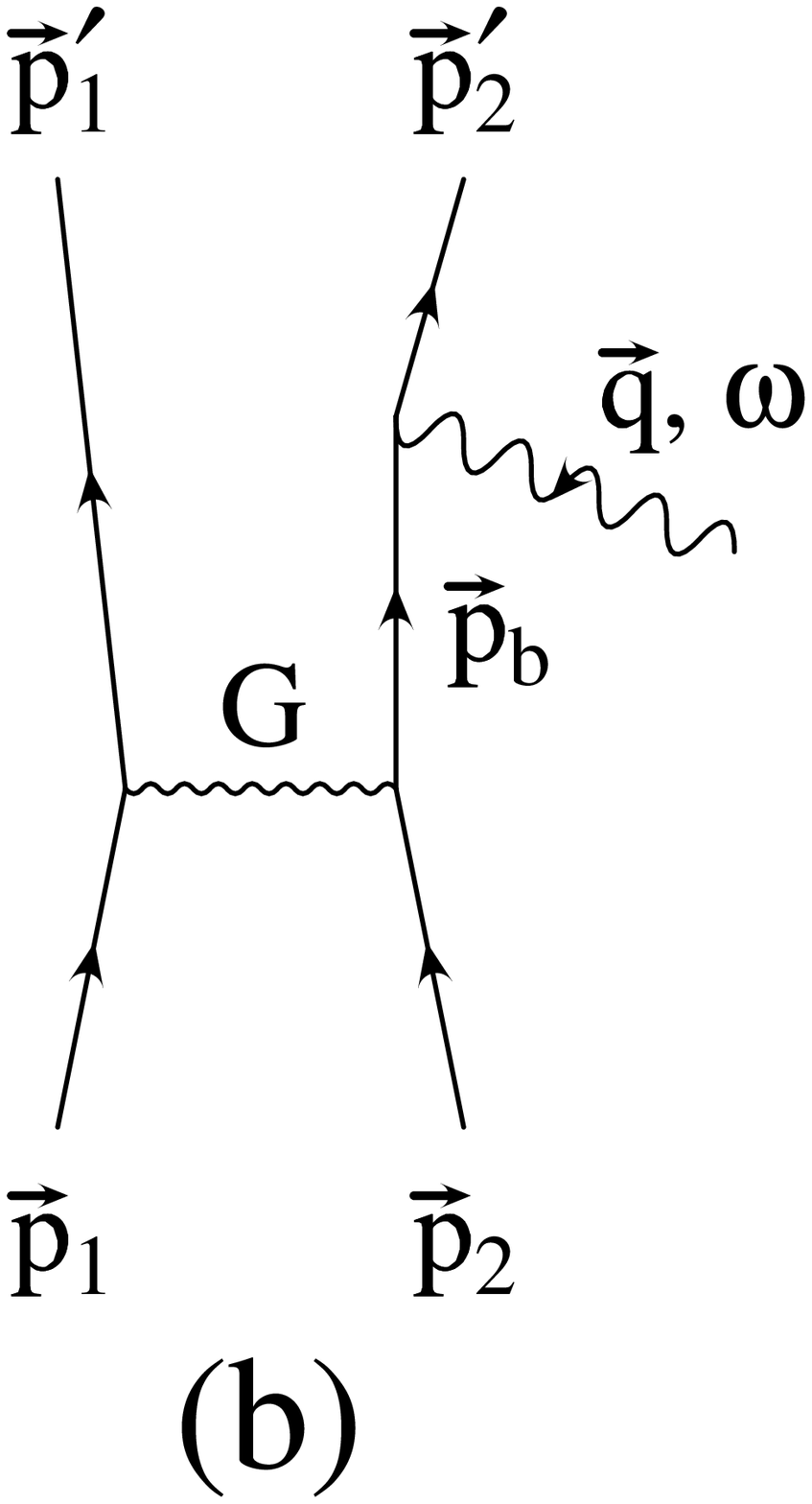,scale=0.25}
\epsfig{file=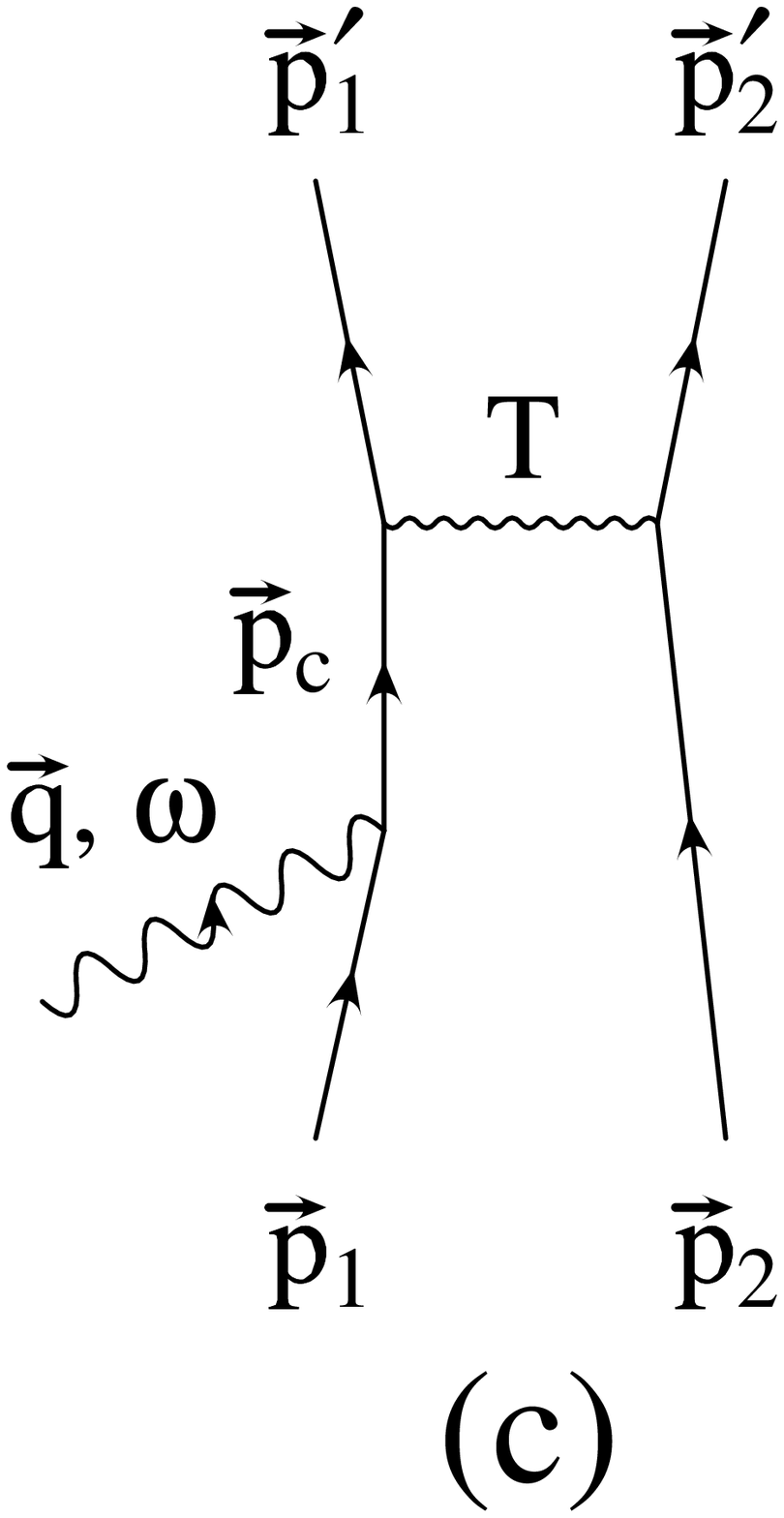,scale=0.25}
\epsfig{file=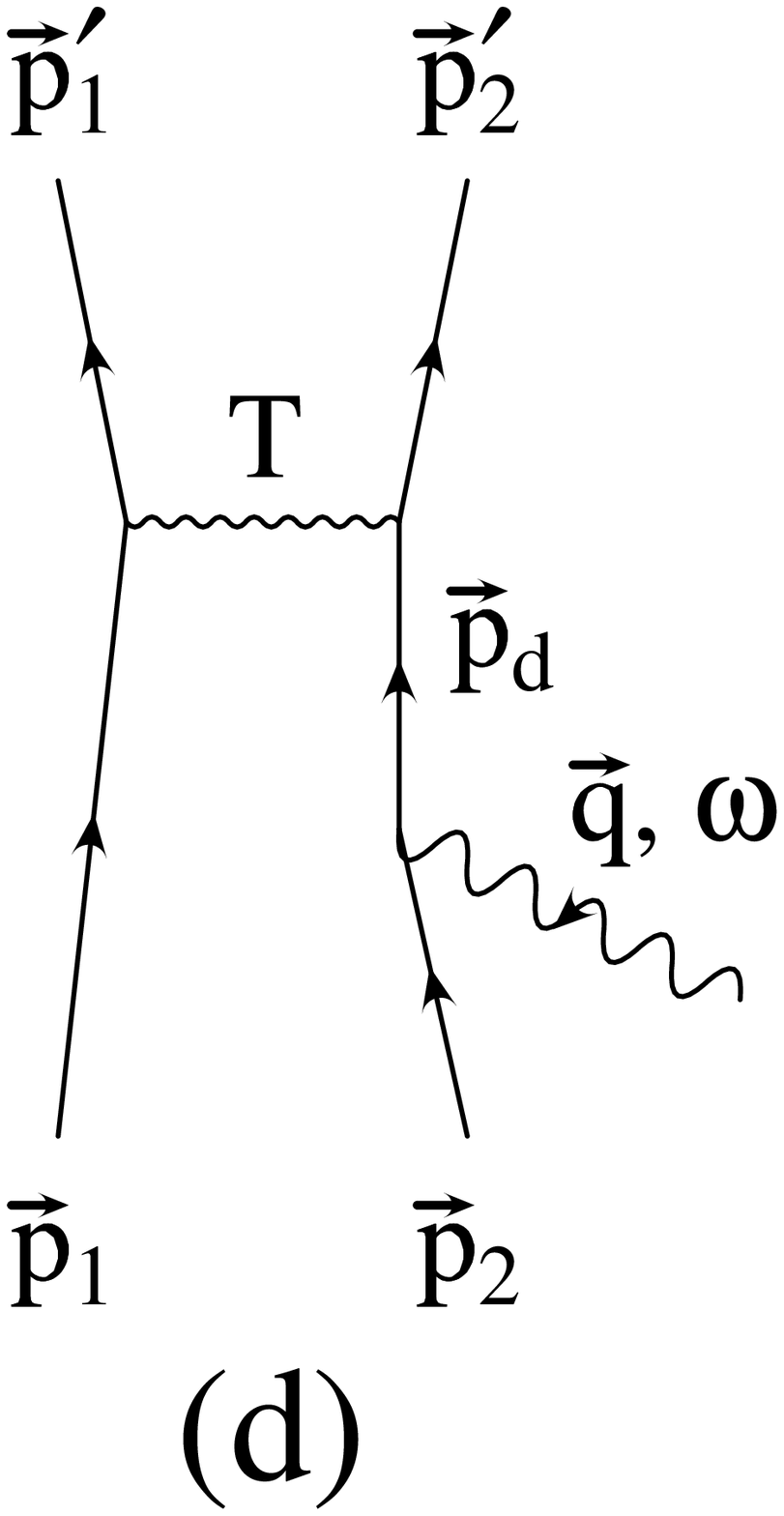,scale=0.25}
\epsfig{file=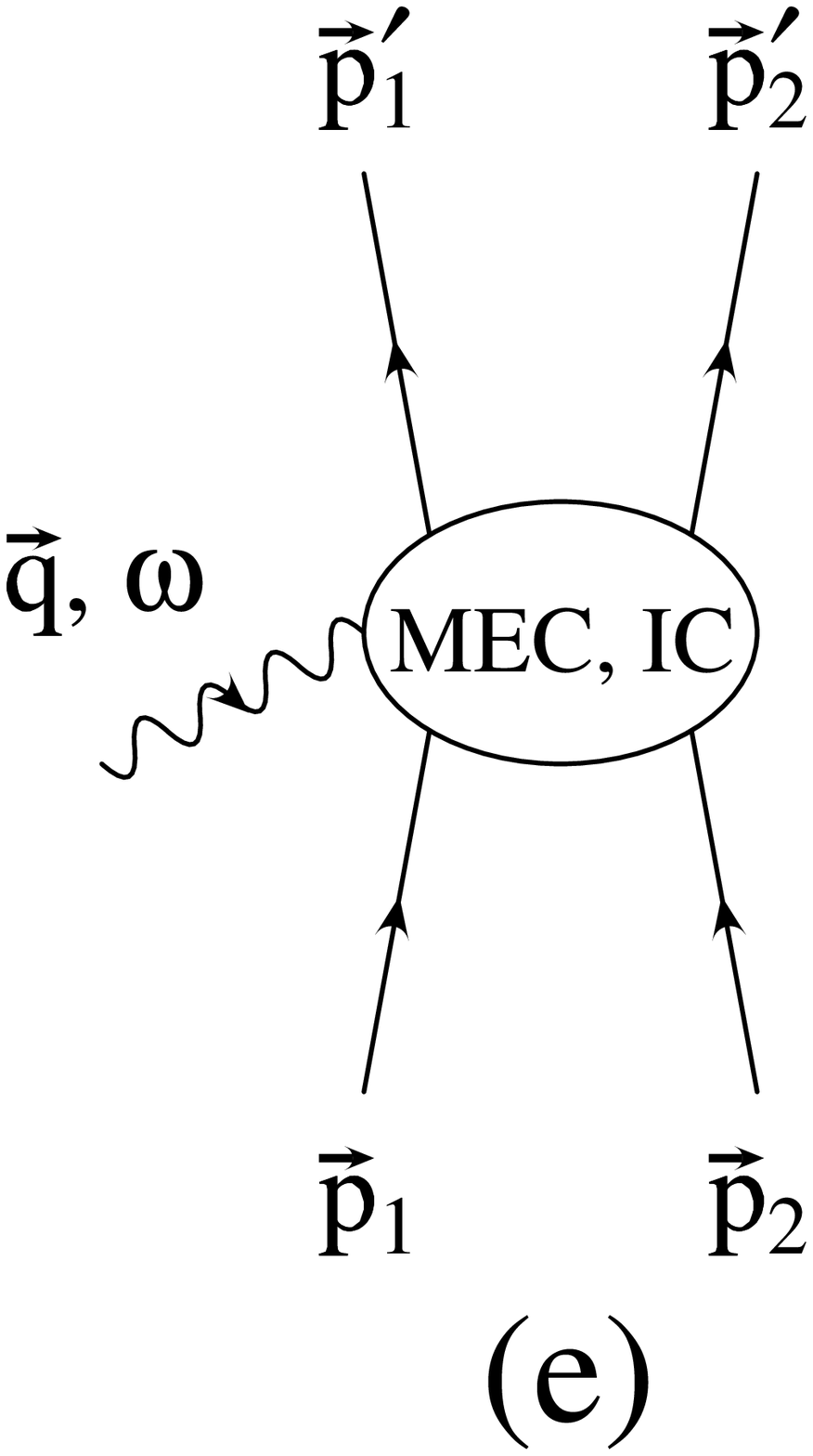,scale=0.25}
\caption{\label{fig1}
Diagrams for the different processes contributing to the $(e,e'2N)$
reaction. Diagram (a) -- (d) show the absorption of the photon by a       
single nucleon. The nucleon-nucleon correlations are described by the $G$
matrix while the re-scattering processes are taken into account in terms
of the $T$ matrix. Diagram (e) depicts photon absorption via meson
exchange (MEC) or isobaric currents (IC).}
\end{center}
\end{figure}

However, correlations in the ground state wave function of the target nucleus
is not the only contribution to the cross section of photoinduced two-nucleon
knock-out. One may also think about processes in which the photon is absorbed
by one nucleon, which propagates off-shell and then shares the absorbed energy
and momentum by scattering on a second nucleon. This process of a final state
interaction (FSI) might be as important as the effect of correlations in the
initial state. In fact, if we describe the absorption of a photon by a
correlated pair of nucleons within Brueckner theory, such processes are
represented by the diagrams (a) and (b) in Fig.~\ref{fig1}. The processes of
the FSI, which we just discussed, can be represented by the diagrams displayed
in (c) and (d) of the same Fig.~\ref{fig1}. Therefore these effects of
correlations and FSI could be interpreted as just two different time orderings
of NN interaction and photon absorption. 

Of course there are some significant differences between these two
contributions. The correlation effect can be described in terms of a Brueckner
G-matrix, which is evaluated at a starting energy of two bound nucleons.
Therefore the resulting G-matrix is real and the corresponding correlated wave
function is bound to heal to the uncorrelated wave function at medium relative
distances. The FSI on the other hand must be described in terms of a solution
of the Bethe-Goldstone equation at a starting energy for the two nucleons above
the Fermi energy. The G-matrix for these starting energies becomes complex and
therefore we denote it by $T$. This implies that the wave function of the two
outgoing nucleons is a scattering wave function which does not show the healing
property. Nevertheless, the symmetry of the contributions displayed in (a) -
(d) of Fig.~\ref{fig1} suggests that one should try to treat them on the same
level of accuracy. This is in fact the main aim of the study presented in this
manuscript.

It should be mentioned that there is another important effect of FSI, which is
usually considered in terms of an optical potential for the outgoing
nucleons\cite{boffi}. This mean field contribution, which accounts for the fact
that the outgoing nucleons will be slowed down, redirected or absorbed in the
field of the residual nucleus, must be be considered in addition to the FSI
contribution discussed above.

Furthermore it is important to keep in mind that there are additional
mechanisms which also contribute to the cross section of two-nucleon
knock-out.   Beside the contributions due to the pair correlations in the ground
state wave function of the target nucleons and the FSI one must also consider
the effects of the two-body terms in the operator for the electromagnetic
current. These two-body terms include meson exchange currents (MEC) which can
be derived from the commutator of the charge density with the nuclear
Hamiltonian to obey the continuity equation. These meson exchange current
contributions should be evaluated in a way consistent with the NN interaction
used to describe correlations and  FSI\cite{mec1,mec2}. In the present work we
include the MEC effects which are due to the exchange of $\pi$ and $\rho$
mesons using the same couplings as in the realistic interactions used to
determine FSI and correlation effects.

In addition, there are contributions to the two-body current which are of
a different origin. Here we, mention the contribution due to the
excitation of intermediate $\Delta$ isobar excitations\cite{ic}. The
contribution of this isobar current (IC) is not constrained by a
continuity equation like the MEC and therefore depends on parameters like
the meson-nucleon $\Delta$ coupling constants and the propagator of the
$\Delta$ in the nuclear medium. 

In the present work we want to evaluate  the nuclear matrix elements for the 
absorption of a photon by a correlated pair of nucleons determining the
effects  of correlations, MEC and FSI in a consistent way from a realistic NN
interaction\cite{machleidt}. As a first step we investigate the absorption of
photons by a pair of nucleons in nuclear matter. For that purpose the
technique, which has been described in \cite{paper1}, accounting for
correlations, MEC and IC effects, is extended to allow the consideration of FSI
as well. The investigation of these various effects shall provide a comparison
of the relative importance of these different mechanisms and their
interference  under various kinematical conditions. 
 
It is of course a serious disadvantage of such a study in nuclear matter
that it does not provide results for a cross section which can directly be
compared with experimental data produced for a specific target nucleus. In
particular, it is impossible to take advantage of the fact that reactions
leading to specific final states of the residual nucleus can be selective
for one of the two-nucleon knock-out mechanisms discussed above. Such a
feature has been observed in theoretical studies of ($e,e'pp$) an ($e,e'pn$)
reactions on $^{16}{\rm O}$\cite{carlot,carlopn}.
On the other hand, however, a study of nuclear matter shall exhibit
general features which are independent on the
specific target nucleus considered and the corresponding long-range or
low-energy correlations. In particular we would like to explore the importance
of the FSI effects as compared to correlation and MEC effects in $pp$ and $pn$
knock-out, if the two nucleons are emitted back to back or in more parallel
directions.

After this introduction we will briefly present the extension of the approach
described in \cite{paper1} to include the effects of final state interaction
between the two ejected nucleons. The discussion of results is presented in
section 3 and the final section 4 contains some concluding remarks.

\section{From the Nuclear Current to the Cross Section}

As it has already often been described in the literature\cite{boffi,wq1,wq2}
the differential cross section for the exclusive
$(e,e'NN)$ reaction can be written
\begin{eqnarray}\label{cross}
\frac{{\rm d}^9\sigma}{{\rm d}\tilde{E_1}{\rm d}\tilde{\Omega_1} {\rm d}
\tilde{E_2}{\rm d}\tilde{\Omega_2} {\rm d}E_e'{\rm d}\Omega_e'} 
&= &\frac{1}{4}\,\frac{1}{(2\pi)^9}\, \tilde{p_1}\, \tilde{p_2}\, \tilde{E_1}\, 
\tilde{E_2}\,
\sigma_{\rm Mott} \,
\Big\{ v_C W_L + v_T W_T +v_S W_{TT} + v_I W_{LT} \Big\}\, 
\nonumber \\ 
&& \times (2\pi)\, \delta (E_f-E_i)
\end{eqnarray}
where the nuclear structure functions $W_i$ ($i=L,T,TT,LT$) contain the
matrix elements of the nuclear current operator which consists of
contributions for the different absorption processes of the virtual photon
carrying momentum $\vec{q}$ and energy $\omega$. Here, the kinematical
variables are the energies $\tilde{E_1},\tilde{E_2}$ and the final momenta
$\tilde{p_1},\tilde{p_2}$ of the two ejected nucleons.

These final nucleon momenta, $\tilde{p_i}$, are different from the 
momenta of the two nucleons after absorption of the photon inside the target, 
which we denote by $p_i'$ because of the mean field effect of the final state
interaction. This mean field contains the attraction of the outgoing nucleons
by the nuclear single-particle potential and essentially reduces the momentum 
of the outgoing nucleon. In infinite nuclear matter this retardation 
can simply be parameterised through an effective mass $m^*$ and we obtain
\begin{equation}\label{eq:effm}
\tilde{E_i} = \frac{\tilde{p_i}^2}{2m} =  \frac{{p_i'}^2}{2m^*} + U\,.
\end{equation}

In \cite{paper1} we investigated the influence of
correlations (diagrams (a) and (b) of Figure \ref{fig1}), meson exchange
currents (MEC, \cite{mec1,mec2}) arising from pions and $\rho\,$ mesons,
and isobaric currents (IC, \cite{ic}) (diagram (e) of Figure \ref{fig1})
on the structure functions of $(e,e'pp)$ and $(e,e'pn)$ reactions. As an example
we recall the expression for the matrix elements entering the structure
functions from the correlation effect displayed in Figure \ref{fig1}(a)
\begin{equation}\label{gmatelement}
\int {\rm d}\vec{p}_a \, \langle\, \vec{p}_1{\!'}\, |\, \vec{\varepsilon}
\cdot\vec{J}_{\gamma NN}\, |\, \vec{p}_a\, \rangle
\, S_2(\vec{p}_a,\vec{p}_2{\!'}) \,
\langle\, \vec{p}_a{\!'}\,\vec{p}_2{\!'}\, |\,
G\, |\, \vec{p}_1\,\vec{p}_2\, \rangle \, .
\end{equation}
Here, $\vec{J}_{\gamma N N}$ is the current operator for the absorption of the
virtual photon with momentum $\vec{q}$ and energy $\omega$ on a single nucleon
\cite{photon}. The matrix element of $G$ is calculated for the starting energy
of two bound nucleons with momenta $\vec p_1$ and $\vec p_2$. Also the
two-particle propagator $S_2$ contains the energy of these bound states and a
Pauli operator which guarantees that the momenta $\vec{p}_a$ and
$\vec{p}_2{\!'}$ are larger than the Fermi momentum. All this ensures that
the dynamical correlation function, which is described in terms of $S_2$ and
$G$, exhibits the so-called healing property.

In close analogy to this correlation term, we construct the
single-particle current contribution  with a re-scattering process
taking place after the absorption of a virtual photon on a single bound
nucleon. In order to give more insight in the evaluation of the corresponding
structure functions entering the cross section in eq.(\ref{cross}), we take
diagram (c) of Figure \ref{fig1} as an example. 

The matrix element for the absorption process displayed in Figure
\ref{fig1}(c) for initial momenta $\vec{p}_1$ and $\vec{p}_2$ and
final momenta $\vec{p}_1{\!'}$ and $\vec{p}_2{\!'}$ of the nucleon-nucleon
pair reads
\begin{equation}\label{tmatelement}
\int {\rm d}\vec{p}_c \, \langle\, \vec{p}_1{\!'}\,\vec{p}_2{\!'}\, |\,
T\, |\, \vec{p}_c\,\vec{p}_2\, \rangle \, S_1(\vec{p}_c) \,
\langle\, \vec{p}_c\, |\, \vec{\varepsilon}\cdot\vec{J}_{\gamma N
N}\, |\, \vec{p}_1\, \rangle.
\end{equation}
The one-nucleon propagator $S_1(\vec{p}_c)$ describes the propagation of the
nucleon after absorption of the energy and momentum of the photon. It also
contains a Pauli operator for the momentum  of the intermediate nucleon
$\vec{p}_c$. Finally, the re-scattering after the
photon absorption is described via the $T$ matrix element $\langle\,
\vec{p}_1{\!'}\,\vec{p}_2{\!'}\, |\, T\, |\, \vec{p}_c\,\vec{p}_2\,
\rangle $.   The $T$ matrix
entering this calculation of the re-scattering process is derived from the
solution of the Bethe-Goldstone equation
for starting energies which lie above the threshold of twice the Fermi
energy. In contrast to the case of eq.(\ref{gmatelement}) 
this leads to complex $T$ matrix elements.

\begin{figure}[tbh]
\epsfig{file=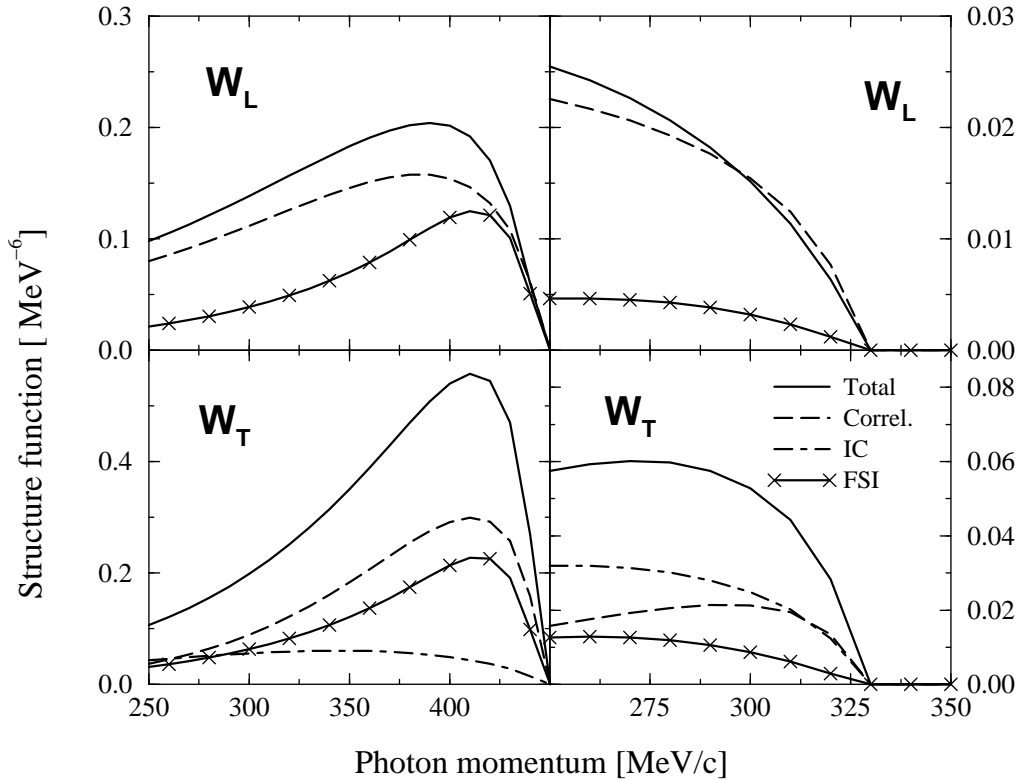,scale=0.7}
\vspace{.5cm}
\caption{\label{fig2}
Longitudinal (upper part) and transverse structure functions (lower part)
for the knock-out of a proton-proton pair in a 'super parallel' kinematical
situation with angles $\theta_{p,1}'=0^{\rm o}$ and
$\theta_{p,2}'=180^{\rm o}$ of the two protons with respect to the
direction of the photon momentum. The left part of the figure assumes
final kinetic energies $T_{p,1}=156\,{\rm MeV}$ and $T_{p,2}=33\,{\rm
MeV}$ of the two protons while in the right part the final kinetic
energies are $T_{p,1}=116\,{\rm MeV}$ and $T_{p,2}=73\,{\rm MeV}$.
The photon energy was chosen to be $\omega=215\,{\rm MeV}$ in all cases.
Together with the total structure functions (solid line) the
contributions arising from correlations (dashed line), FSI (solid line
with crosses), and IC (dot-dashed line) are shown. Please note the
different scales in the various parts of the figure. All structure functions
have been multiplied by a common factor $10^{10}$.}
\end{figure}

\section{Results and Discussion}

The calculations are performed in nuclear matter at saturation density
($k_F=1.35\,{\rm fm}^{-1}$). The coupling constants for the $\pi$ exchange
are $f_{\pi N N}=1.005$, $f_{\pi N \Delta}=2f_{\pi N N}$, and $f_{\gamma N
\Delta}=0.12$. For the $\rho$ exchange, the coupling constants were chosen
to be $g_{\rho}^2/4\pi=0.86$, $\kappa_{\rho}=6.1$. The cutoff masses for
the $\pi$-nucleon and $\rho\,$-nucleon form factors were fixed at
$\Lambda_{\pi}=1.3\,{\rm GeV}$ and $\Lambda_{\rho}=1.95\,{\rm GeV}$
respectively. The $\pi$-nucleon and the $\rho$-nucleon vertices are
compatible with the $\pi$ and the $\rho$ exchange part of the
One-Boson-Exchange potential BONN A \cite{machleidt} we used to determine
the $G$ and the $T$ matrix for calculating the effects of NN correlations
and re-scattering processes, respectively. The same potential has also been
used to determine the single-particle energies for the nucleons in the
nuclear medium which were parametrised according to (\ref{eq:effm}) by
$m^*=623\, {\rm MeV}$ and $U=-86.8\, {\rm MeV}$.

For all presented structure functions, the MEC contributions contain the
seagull and meson in-flight currents for the exchange of a $\pi$ meson as well
as the seagull, pair and in-flight currents for the $\rho$ meson. As we
reported already in \cite{paper1}, MEC contributions like the photon
absorption on a pair of $\pi$ and $\rho$ mesons were found to give
negligible contributions in all kinematical setups we investigated.

\begin{figure}[tbh]
\begin{center}
\epsfig{file=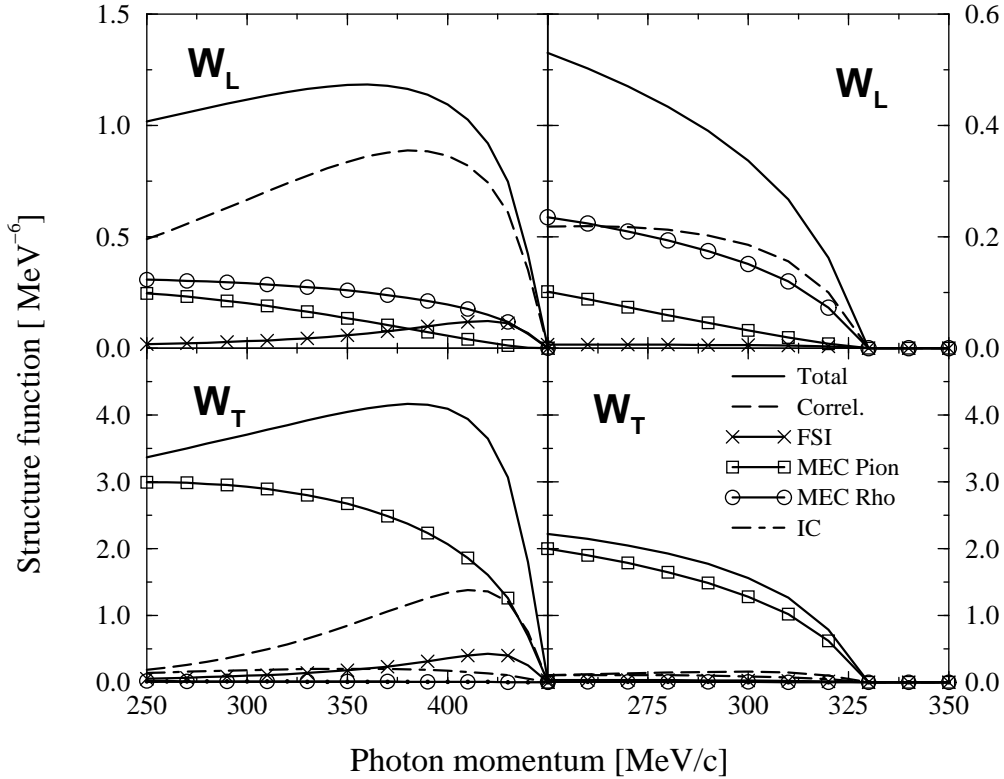,scale=0.7}
\end{center}
\vspace{.5cm}
\caption{\label{fig3}
Longitudinal (upper part) and transverse structure functions (lower part)
for the knock-out of a proton-neutron pair in a 'super parallel'    
kinematical situation with angles $\theta_{p}'=0^{\rm o}$ and
$\theta_{n}'=180^{\rm o}$ of the proton and the neutron with respect to  
the direction of the photon momentum. The left part of the figure assumes
final kinetic energies $T_{p}=156\,{\rm MeV}$ and $T_{n}=33\,{\rm MeV}$ of
the two protons while in the right part the final kinetic energies are
$T_{p}=116\,{\rm MeV}$ and $T_{n}=73\,{\rm MeV}$. The photon energy was
chosen to be $\omega=215\,{\rm MeV}$ in all cases. Together with the total
structure functions (solid line) the contributions arising from
correlations (dashed line), FSI (solid line with crosses), MEC (solid line
with squares for the $\pi$ exchange, solid lines with circles for the
$\rho$ exchange), and IC (dot-dashed line) are shown. Please note the
different scales in the various parts of the figure. All structure functions
have been multiplied by a common factor $10^{10}$.}
\end{figure}

Our earlier investigations\cite{paper1} showed that the so-called 'super 
parallel' kinematical setup\cite{grab1} is very appropriate to investigate 
effects of nucleon-nucleon correlations. Therefore we will first present 
results of calculations including the re-scattering processes
for this kinematical situation. For that purpose we chose once again the setup
in which one of the two ejected nucleons moves in direction of the
momentum of the absorbed photon and the other antiparallel to this
direction. In Figures \ref{fig2} and \ref{fig3} we show the longitudinal
and transverse structure functions for the knock-out of a proton-proton
and a proton-neutron pair, respectively. In both cases, the virtual photon
carries an energy of $\omega=215\,{\rm MeV}$. 

For the structure functions of the $(e,e'pp)$ reaction displayed in Figure
\ref{fig2} we find non-negligible contributions from the FSI
re-scattering term. This is true in particular in the case of a very asymmetric 
splitting of the available energy to the the kinetic energy of the outgoing 
protons ($T_{p,1}=156\,{\rm MeV}$ and $T_{p,2}=33\,{\rm MeV}$), which is
presented in the left part of the Figure. The contribution from FSI 
re-scattering is almost as large as the correlation effects. Both are
significantly stronger than the contribution of the isobar current. Correlation
and FSI effects add up in a rather coherent way to the total cross section. 

The values for the structure functions are reduced by almost one order of
magnitude if one considers the more symmetric distribution of the available
energy among the two emitted protons, which is considered in the right part of
the figure ($T_{p,1}=116\,{\rm MeV}$ and $T_{p,2}=73\,{\rm MeV}$). This shows
that correlations as well as FSI effects are not very efficient in
redistributing energy and momentum from the proton emerging in direction
parallel to the momentum transfer $\vec q$ to the second proton,  which is
emitted antiparallel to $\vec q$. In this example the isobar current yields the
largest contribution to the transverse structure function. It is worth noting
that the FSI re-scattering effect is smaller than the correlation effect. This
is true in particular for the longitudinal structure function.

In the case of the knock-out of a proton-neutron pair in Figure \ref{fig3},
the effects of correlations and final state interaction are masked to a large
extent by the effects of meson exchange currents (MEC). The large MEC
contribution is one reason for the fact that the structure functions of 
proton-neutron knock-out are almost one order of magnitude larger than the
corresponding structure functions for proton-proton emission. In the case of
the asymmetric energy splitting (large kinetic energy for the proton emitted
parallel to $\vec q$), however the dominant contribution to the longitudinal
structure function originates from correlation effects. The larger correlation
contribution in $pn$ knock-out as compared to $pp$ reflects the importance of
tensor correlations in the case of the $pn$ pair.  In contrast to the
proton-proton knock-out, the influence of the final state interaction is
negligible for the more symmetric distribution of the kinetic energies (see
right part of Figure \ref{fig3}).

Comparing the absolute values of the contributions arising from
re-scattering processes, one finds that these contributions
are about twice as large as for the knock-out of a proton-proton pair. This
demonstrates that the proton-neutron interaction is in general stronger than the
proton-proton interaction. The enhancement of the correlation effect in going
from the $pp$ case to the $pn$ case, however, is much stronger.

\begin{figure}[tbh]
\begin{center}
\epsfig{file=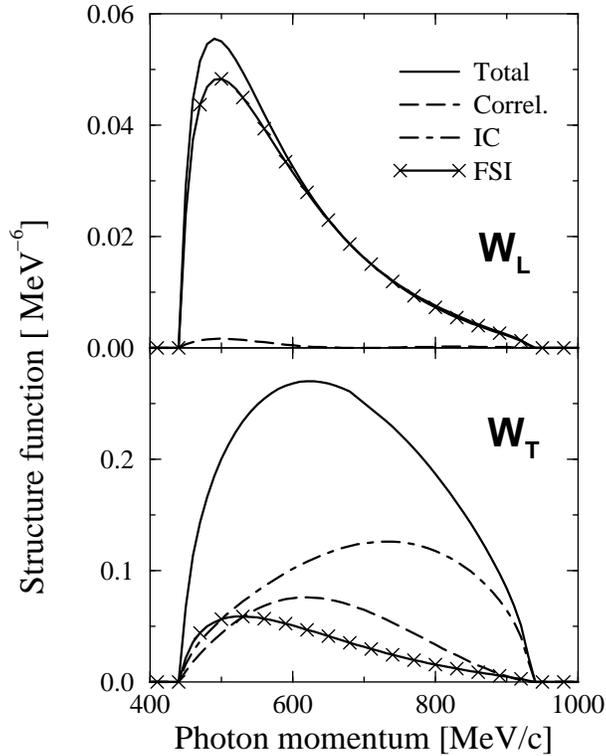,scale=0.7}
\end{center}
\vspace{.5cm}
\caption{\label{fig4}
Longitudinal (above) and transverse structure functions (below) for the
knock-out of a proton-proton pair. The angles of the outgoing protons are
$\theta_{p,1}'=\theta_{p,2}'=30^{\rm o}$ while the final
kinetic energies are $T_{p,1}=T_{p,2}=70\,{\rm MeV}$. The photon energy
was chosen to be $\omega=230\,{\rm MeV}$. The total structure functions
(solid line) consist of the one-body current (dashed line for
correlations, solid line with crosses for FSI) and the IC contribution
(dot-dashed line). All structure functions
have been multiplied by a common factor $10^{10}$.}
\end{figure}

\begin{figure}[tbh]
\begin{center}
\epsfig{file=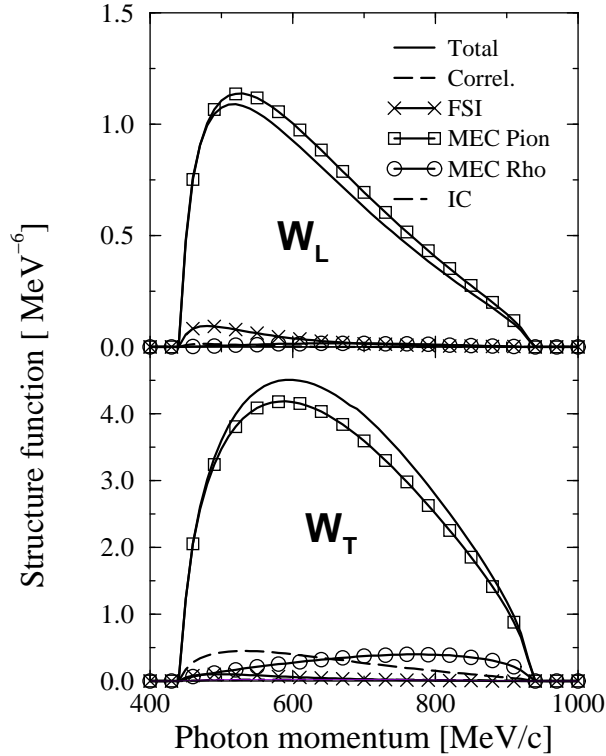,scale=0.7}
\end{center}
\vspace{.5cm}
\caption{\label{fig5}
Longitudinal (above) and transverse structure functions (below) for the
knock-out of a proton-neutron pair. The angles of the outgoing nucleons are
$\theta_{p}'=\theta_{n}'=30^{\rm o}$ while the final kinetic energies are
$T_{p}=T_{n}=70\,{\rm MeV}$. The photon energy was  chosen to be
$\omega=230\,{\rm MeV}$. The total structure functions (solid line) consist of
the one-body current (dashed line for correlations, solid line with crosses for
FSI), the MEC contributions (solid lines with squares and circles) and the IC
contribution (dot-dashed line). All structure functions have been multiplied by
a common factor $10^{10}$.}
\end{figure}

Sometimes the argument has been used that the FSI re-scattering effects should
be minimal if the two nucleons are emitted back to back, the case which we have
considered up to now. It is evident that FSI effects are very important if the
two nucleons are emitted parallel or even form a deuteron as in ($e,e'd$)
experiments. How important is the FSI re-scattering contribution in
kinematical setups, which are in between antiparallel and parallel emission? To
answer this question we consider a kinematical set up, which we also inspected
in our earlier investigations of \cite{paper1}. Here, the two nucleons are
ejected in a symmetrical way ($\theta_{p,1/p}'=\theta_{p,2/n}'=30^{\rm o}$, on
opposite sides of the momentum transfer $\vec q$) having the same kinetic
energies of $70\,{\rm MeV}$. Results for the structure functions of $pp$ and
$pn$ knock-out are displayed in Figures \ref{fig4} and \ref{fig5}, respectively.

In the longitudinal channel of the $(e,e'pp)$
reaction (Figure \ref{fig4}) the contribution from final state interaction 
is clearly the dominating one. Contributions from correlations are only of 
minor importance. For the transverse channel, however, both effects yield
contributions of similar size, which are furthermore of the same order of
magnitude as the isobar contribution. 

The absolute strength of re-scattering contribution to $(e,e'pn)$ displayed in
Figure \ref{fig5} is about twice as large as for the corresponding ($e,e'pp$)
case. The cross section for $pn$ emission in this setup, however, is by far
dominated by the meson exchange currents which are related to the pion exchange.

\section{Summary and Conclusions}

The effects of re-scattering processes in the final state interaction of
two-nucleon knock-out reactions induced by electron scattering have been
investigated. The corresponding matrix elements leading to longitudinal and
transverse nuclear structure functions have been calculated for a pair of
nucleons in nuclear matter. The contributions from re-scattering are calculated
in a way which is consistent with the evaluation of correlation effects and
meson exchange current contributions. All contributions are derived from the
same realistic meson exchange model of the NN interaction. 
This allows a systematic study of the relative importance of these effects 
under various kinematical conditions.

The effects of final state interactions (FSI) are very important if the two
nucleons are emitted in directions with small angle in between. As an example
we consider the longitudinal structure function for $(e,e'pp)$ with an angle of
$60^{\rm o}$, between the momenta of the outgoing protons. This example is
completely dominated by the FSI effects. FSI contributions are larger by a
roughly a factor of two in ($e,e'pn$) as compared to ($e,e'pp$) processes.
Since other mechanisms, like correlation effects and meson exchange current
contributions, are enhanced by a factor larger than two in $pn$ as compared to
$pp$ knock-out, the relative importance of  FSI are less pronounced in
($e,e'pn$) reactions.  

FSI effects are non-negligible also in the case of back to back
emission of the two nucleons. In this case the FSI effects tend to be smaller
than correlation effects. Nevertheless they yield contributions of similar size
and therefore should be treated in a way which is consistent with the treatment
of correlation effects.   

This research project has been supported by the SFB 382 of the
Deutsche Forschungsgemeinschaft.

\end{document}